\documentclass[10pt,a4paper]{article}
\usepackage[utf8x]{inputenc}
\usepackage{polski}
\usepackage{amsmath}
\usepackage{amsthm}
\usepackage{amssymb}

\usepackage{algorithmic}

\usepackage{float}

\newcommand{\TOA}{\textbf{to }}

\newcommand{\simf}{\mathrm{sim}}
\newcommand{\maxlen}{\mathrm{maxlen}}
\newcommand{\simlen}{\mathrm{simlen}}
\newcommand{\simlenmax}{\mathrm{simlenmax}} 
\newcommand{\scf}{\mathrm{sc}} 

\newfloat{program}{tbp}{lop}
\floatname{program}{Algorytm}

\newtheorem{lemat}{Lemat}

\title{Efektywny algorytm wyznaczający miarę podobieństwa wyrazów za pomocą uogólnionej metody $n$-gramów (miary Niewiadomskiego)}
\author{prof. dr hab. Stanisław Goldstein \\ Uniwersytet Łódzki \and mgr inż. Piotr Beling \\ Uniwersytet Łódzki }	
\date{\today}
\begin{document}

\maketitle

\begin{abstract}
W artykule przedstawiono efektywny (dynamiczny) algorytm wyznaczający miarę podobieństwa wyrazów za pomocą uogólnionej metody $n$-gramów (miary Niewiadomskiego \cite{NiewiadomskiDR}). Uzasadniono także poprawność działania algorytmu i~oszacowano jego złożoność obliczeniową.
\end{abstract}

\renewcommand{\abstractname}{Abstract}
\begin{abstract}
This article presents effective (dynamic) algorithm for solving a problem of counting the number of substrings of given string which are also substrings of second string. Presented algorithm can be used for example for quick calculation of strings similarity measure using generalized $n$-gram method (Niewiadomski measure \cite{NiewiadomskiDR}), which are shown. Correctness and complexity analyses are included.
\end{abstract}

\section{Oznaczenia}
Jeśli przez $w$, $s$ będą oznaczone słowa, to:
\begin{itemize}
	\item $|w|$ - długość słowa $w$ (liczba liter),
	\item litery w słowie $w$ mają indeksy od $0$ do $|w|-1$,
	\item $w[a..b]$ (dla $a, b \in \mathbb{N}$) - podsłowo słowa $w$ od litery $a$-tej do $b$-tej (włącznie) lub słowo puste gdy $a>b$,
	\item $w[a..b)$ (dla $a, b \in \mathbb{N}$) - podsłowo $w[a..b-1]$ słowa $w$,
	\item $w[a..)$ (dla $a \in \mathbb{N}$) - podsłowo $w[a..|a|)$ (sufiks słowa $w$ rozpoczynający się od $a$-tej litery),
	\item $w[a]$ (dla $a \in \mathbb{N}$) $a$-ta litera słowa $w$ ($w[a]=w[a..a]$),
	\item $ws$ - konkatenacja słów $w$ i $s$, tj. słowo długości $|w|+|s|$, takie, że $(ws)[0..|w|)=w$ i $(ws)[|w|..)=s$;
	\item zachodzi $w=w[0..|w|-1]=w[0..|w|)=w[0..)=w[0]w[1]\ldots w[|w|-1]$.
\end{itemize}

\section{Metoda $n$-gramów}
Metody $n$-gramów ($1$-gramów czyli unigramów, $2$-gramów czyli bigramów, $3$-gramów czyli trigramów, itd.) są popularnymi technikami analizy tekstów o~szerokich możliwościach zastosowania, por. np. \cite{use1, use2, use3, use4, use5, wikipedia}.

Miary $n$-gramów bazują na zliczaniu $n$-literowych słów (zwanych $n$-gramami) będących równocześnie podsłowami obu porównywanych słów.
Im więcej wspólnych $n$-gramów mają porównywane słowa tym bardziej są do siebie podobne.
Więcej szczegułów można znaleźć w artykule \cite{Kondrak}.

%

\section{Uogólniona miara $n$-gramów}
Uogólnienie miar $n$-gramów polegające na zliczaniu podsłów dowolnej długości zaproponował A.~Niewiadomski w~rozprawie \cite{NiewiadomskiDR}.

Podobieństwo słów $w_1$ i $w_2$ wg. uogólnionej miary $n$-gramów wynosi $\min(\simf(w_1, w_2), \simf(w_2, w_1))$,
przy czym funkcja $\simf$ zdefiniowana jest następująco:
\begin{equation}\label{def:ngram}
\simf(w_1, w_2)=\frac{2}{N^2+N}\sum_{i=1}^{\min(|w_1|, |w_2|)}\sum_{j=0}^{|w_1|-i}h(i,j)
\end{equation}
gdzie:
\begin{itemize}
	\item $N=\max(|w_1|, |w_2|)$
	\item $h(i, j)=\begin{cases} 1, & \mbox{gdy }w_1[j..j+i)\mbox{ jest podsłowem } w_{2}\\ 0, & \mbox{w przeciwnym razie} \end{cases}$
\end{itemize}
Funkcja $\simf$ liczy iloraz:
\begin{itemize}
	\item liczbę (niekoniecznie różnych) podsłów $w_1$ będących równocześnie podsłowami $w_2$, przez:
	\item liczbę wszystkich (niekoniecznie różnych) podsłów dłuższego ze słów: $w_1$, $w_2$ (jest ich $\frac{N^2+N}{2}$).
\end{itemize}

Funkcja $\simf$ nie jest symetryczna.

Miara Niewiadomskiego może być zastosowana do porównywania dowolnych ciągów liter, zarówno całych dokumentów tekstowych, całych zdań, fraz lub pojedynczych wyrazów.
Jej przydatność w~praktyce potwierdził m.in. J.~Adamczak w~pracy \cite{AdamczakMGR}. Wykorzystał on ją w~budowie systemu ekspertowego wspomagającego diagnostykę medyczną.
Zadaniem jego opartego o~logikę rozmytą, regułowego systemu było porównanie (sklasyfikowanie) nowego, opisanego stanem pacjenta przypadku z~zawartymi w~bazie zdiagnozowanymi przypadkami oraz podanie  zbioru możliwych diagnoz.

\section{Algorytm naiwny i jego złożoność}
Oznaczmy: $a=|w_1|$, $b=|w_2|$.
Algorytm naiwny\footnote{Sprawdzający (dla kolejnych długości $i=1, 2, \ldots, \min(a, b)$) kolejno wszystkie podsłowa $i$-literowe słowa $w_1$ (jest ich $a-i+1$) czy są podsłowami $w_2$. Jest $b-i+1$ $i$-literowych podsłów $w_2$. Porównanie dwóch słów długości $i$ może (w~razie równości) wymagać $i$ porównań symboli.}\footnote{użyty m.in. w~implementacji systemu ekspertowego przez J.~Adamczaka \cite{AdamczakMGR}} wyznaczający wartość funkcji $\simf$ wprost ze wzoru \ref{def:ngram} wykona (w~pesymistycznym przypadku) następującą ilość porównań symboli:
\[
\sum_{i=1}^{\min(a,b)}(a-i+1)(b-i+1)i
\]
Ze względu na symetryczność powyższego wzoru, możemy, bez straty ogólności obliczeń założyć, że $\min(a, b)=a$, mamy wtedy:
\begin{align*}
& \sum_{i=0}^{a-1}(a-i)(b-i)(i+1) = \sum_{i=0}^{a-1}(ab-ai-bi+i^2)(i+1) = \\
& ab\sum_{i=0}^{a-1}(i+1)-(a+b)\sum_{i=0}^{a-1}i(i+1)+\sum_{i=0}^{a-1}i^2(i+1) =	\\
& ab\sum_{i=1}^{a}i - (a+b)\sum_{i=1}^{a}(i^2-i) + \sum_{i=0}^{a-1}i^3 + \sum_{i=0}^{a-1}i^2 = \\
& \frac{aba(a+1)}{2} - (a+b)\sum_{i=1}^{a}i^2 + (a+b)\sum_{i=1}^{a}i + (\frac{(a-1)a}{2})^2 + \sum_{i=1}^{a}i^2 - a^2 = \\
& \frac{a^3b+a^2b}{2} - (a+b+1)\frac{a(a+1)(2a+1)}{6} + \frac{(a+b)a(a+1)}{2} +(\frac{a^2-a}{2})^2 - a^2 = \\
& \frac{1}{12}(2a^3b + 6a^2b - a^4 - 10a^3 - 11a^2 - 8ab - 2a) =  \Theta(a^3b)
\end{align*}
Stąd, złożoność czasowa algorytmu naiwnego wynosi $\Theta(|w_1||w_2|\min^2(|w_1|, |w_2|))$. Jego złożoność pamięciowa to $\Theta(1)$.

Algorytm naiwny można poprawić używając w jego konstrukcji algorytmu wyszukującego wzorca w~czasie liniowym (np. Knutha-Morrisa-Pratta). Otrzymamy wtedy algorytm o~złożoności czasowej $\Theta(|w_1||w_2|\min(|w_1|, |w_2|))$.

\section{Algorytm efektywny i jego złożoność}
Oznaczmy: $\scf(w_1, w_2)=\sum_{i=1}^{\min(|w_1|, |w_2|)}\sum_{j=0}^{|w_1|-i}h(i,j)$ (zachodzi: $\simf(w_1, w_2)=\frac{2}{N^2+N}\scf(w_1, w_2)$). $\scf(w_1, w_2)$ to ilość (niekoniecznie różnych) podsłów $w_1$ będących równocześnie podsłowami $w_2$.

\begin{lemat}[o funkcji h]\label{lemath}
Jeśli $h(i, j)=1$ to dla dowolnego $y<i$: $h(y, j)=1$.
\end{lemat}

\begin{proof}
Jeśli $h(i, j)=1$ to $w_1[j..j+i)$ jest podsłowem $w_{2}$, tzn. istnieje w~$w_{2}$ pozycja $x$ taka, że
$w_1[j..j+i)=w_{2}[x..x+i)$, stąd zaś $w_1[j..j+i-1)=w_{2}[x..x+i-1)$ (czyli h(i-1, j)=1), $w_1[j..j+i-2)=w_{2}[x..x+i-2)$ (czyli h(i-2, j)=1), itd.
\end{proof}

Z lematu \ref{lemath} wynika, że dla dowolnego $j$ istnieje $i\in\{0, 1, \ldots\}$, takie, że:
\begin{itemize}
	\item $h(y, j)=1$ dla dowolnego $y \leq i$,
	\item $h(y, j)=0$ dla dowolnego $y > i$.
\end{itemize}

Inaczej mówiąc, dla danego indeksu $j$ początku podsłowa $w_1$ można znaleźć największą jego długość $i$, taką że:
\begin{itemize}
	\item $w_1[j..j], w_1[j..j+1], \ldots, w_1[j..j+i)$ są podsłowami $w_2$,
	\item $w_1[j..j+i], w_1[j..j+i+1], \ldots, w_1[j..|w_1|)$ nie są podsłowami $w_2$.
\end{itemize}
Oznaczmy tą liczbę symbolem $\maxlen(w_1, w_2, j)$.

Można policzyć liczbę $\scf(w_1, w_2)$ jako sumę:
\begin{equation}\label{scmaxlensum}
\scf(w_1, w_2)=\sum_{j=0}^{|w_1|-1}\maxlen(w_1, w_2, j)
\end{equation}

Rozważmy kolejne sufiksy słowa $w_1$: $p_1=w_1[|w_1|-1..), p_2=w_1[|w_1|-2..), \ldots, p_i=w_1[|w_1|-i..)$. Zachodzi $|p_i|=i$.

Niech $\simlen$ będzie funkcją taką, że $\simlen(w_1, w_2, n, l)=x$ wtedy i tylko wtedy, gdy:
$p_n[0..x)=w_2[l..l+x)$ oraz (w przypadku gdy $x \neq n$) $p_n[x] \neq w_2[l+x]$.

Czyli $\simlen(w_1, w_2, n, l)=x$ gdy słowa $p_n$ i $w_2[l..)$ mają dokładnie $x$ pierwszych takich samych liter (tj. ich najdłuższy wspólny prefiks ma~długości $x$).

Przyjmijmy dodatkowo $\simlen(w_1, w_2, 0, l)=0$ dla dowolnego $l$.

\begin{lemat}\label{fastsimlenfound}
Dla $n>0$:
\begin{itemize}
	\item $\simlen(w_1, w_2, n, l)=0$ gdy $p_n[0] \neq w_2[l]$
	\item $\simlen(w_1, w_2, n, l)=1+\simlen(w_1, w_2, n-1, l+1)$ gdy $p_n[0] = w_2[l]$
\end{itemize}
\end{lemat}
\begin{proof}
Pierwsza równość wynika wprost z~definicji $\simlen$.
Druga także z faktów, że $p_n=p_n[0]p_{n-1}$ i $w_2[l..)=w_2[l]w_2[l+1..)$.
\end{proof}

Oznaczmy:
\begin{equation}
\simlenmax(w_1, w_2, n)=\max(\simlen(w_1, w_2, n, 0), \ldots, \simlen(w_1, w_2, n, |w_2|-1))
\end{equation}
Łatwo zauważyć, że $\simlenmax(w_1, w_2, n)=x$ oznacza, że najdłuższy prefiks słowa $p_n$ będący podsłowem $w_2$ jest długości $x$, tj. że $p_n[0..x-1]$ ($=w_1[|w_1|-n..|w_1|-n+x)$) jest podsłowem $w_2$ i~równocześnie (gdy $x \neq |p_n|$) nie jest nim $p_n[0..x]$ ($=w_1[|w_1|-n..|w_1|-n+x]$).

Jeśli więc $\simlenmax(w_1, w_2, n)=x$ to $h(x, |w_1|-n)=1$ i~równocześnie $h(x+1, |w_1|-n)=0$.

Z lematu \ref{lemath} także: $h(x-1, |w_1|-n)=1, h(x-2, |w_1|-n)=1, \ldots, h(1, |w_1|-n)=1$.

Zachodzi więc $\simlenmax(w_1, w_2, n)=\maxlen(w_1, w_2, |w_1|-n)$ i (na~podstawie \ref{scmaxlensum}):
\begin{equation}\label{scsimlenmaxsum}
\scf(w_1, w_2)=\sum_{j=0}^{|w_1|-1}\simlenmax(w_1, w_2, j)
\end{equation}

\begin{program}
\caption{dynamiczny algorytm wyznaczający wartość \ensuremath{\scf(w_1, w_2)}}
\label{alg:dyn:sc}
\begin{algorithmic}
\STATE \COMMENT{funkcja \ensuremath{\scf(w_1, w_2)}}
\STATE \COMMENT{tablica liczb simlen ma długość \ensuremath{|w_2|+1}}
\FOR{$b = 0$ \TOA $|w_2|$}
	\STATE $simlen[b] \gets 0$
\ENDFOR
\STATE $result \gets 0$
\STATE $alast \gets |w_1|$
\WHILE {$alast \neq 0$}
	\STATE $alast \gets alast-1$
	\STATE $max \gets 0$
	\STATE $s \gets 0$
	\FOR {$b = 0$ \TOA $|w_2|-1$}
		\IF {$w_2[b]=w_1[alast]$}
			\STATE $simlen[s] \gets simlen[s+1]+1$
			\IF {$simlen[s]>max$}
				\STATE $max \gets simlen[s]$
			\ENDIF
		\ELSE
			\STATE $simlen[s] \gets 0$
		\ENDIF
		\STATE $s \gets s+1$
	\ENDFOR
	\STATE $result \gets result+max$
\ENDWHILE
\RETURN $result$
\end{algorithmic}
\end{program}

Algorytm \ref{alg:dyn:sc} wyznacza wartość $\scf$ jako sumę $\simlenmax(w_1, w_2, n)$ dla kolejnych $n=1, 2, \ldots, |w_1|$ (zgodnie z~\ref{scsimlenmaxsum}). Czyni to wykonując $|w_2|$ porównań (dla każdego $n$) korzystając z~programowania dynamicznego i~zależności \ref{fastsimlenfound}. Przy obliczaniu $\simlen(w_1, w_2, n, ...)$ korzysta z~$|w_2|$~wcześniej zapamiętanych wartości $\simlen(w_1, w_2, n-1, ...)$, co wymaga użycia $\Theta(|w_2|)$ dodatkowej pamięci.

\begin{program}
\caption{algorytm wyznaczający miarę podobieństwa słów \ensuremath{w_1} i \ensuremath{w_2} uogólnioną metodą $n$-gramów, tj. liczący wartość \ensuremath{\min(\simf(w_1, w_2), \simf(w_2, w_1))}}
\label{alg:dyn:sim}
\begin{algorithmic}
	\STATE $N \gets \max(|w_1|, |w_2|)$
	\RETURN $2.0 * \min(\scf(w_1, w_2), \scf(w_2, w_1)) / (N * N + N)$
\end{algorithmic}
\end{program}

Algorytmy \ref{alg:dyn:sc} oraz \ref{alg:dyn:sim} mają złożoność czasową $\Theta(|w_1||w_2|)$ (wykonują one odpowiednio $|w_1||w_2|$ i $2|w_1||w_2|$ porównań symboli).
Pamięciowa złożoność algorytmu \ref{alg:dyn:sim} to $\Theta(|w_1| + |w_2|)$.

\end{document}